\documentclass[paper]{JHEP3}

\usepackage{epsfig,multicol,multirow}
\usepackage{amsmath,subfigure,latexsym,amssymb}

\def\lsi{\raise0.3ex\hbox{$<$\kern-0.75em\raise-1.1ex\hbox{$\sim$}}}
\def\gsi{\raise0.3ex\hbox{$>$\kern-0.75em\raise-1.1ex\hbox{$\sim$}}}

\newcommand\fverb{\setbox\pippobox=\hbox\bgroup\verb}
\newcommand\fverbdo{\egroup\medskip\noindent%
                        \fbox{\unhbox\pippobox}\ }
\newcommand\fverbit{\egroup\item[\fbox{\unhbox\pippobox}]}
\newbox\pippobox
\newcommand{\beq}{\begin{equation}}
\newcommand{\eeq}{\end{equation}}
\newcommand{\beqa}{\begin{eqnarray}}
\newcommand{\eeqa}{\end{eqnarray}}

\newcommand{\R}{\mathbb{R}}
\newcommand{\ee}{{\rm e}}

\newcommand {\tr}{{\rm tr\,}}

\preprint{
SAGA-HE-223 \\ 
KEK-TH-1085 \\
UTHEP-514}
\title{
Probability distribution of the index\\
in gauge theory on 2d non-commutative geometry
}

\author{
Hajime Aoki${}^{a}$,
Jun Nishimura${}^{bc}$ and
Yoshiaki Susaki${}^{bd}$\\ 
\llap{$^a$}Department of Physics, Saga University, 
Saga 840-8502, Japan \\
\llap{$^b$}High Energy Accelerator Research Organization (KEK), \\
Tsukuba, Ibaraki, 305-0801, Japan \\
\llap{$^c$}Department of Particle and Nuclear Physics,\\
Graduate University for Advanced Studies (SOKENDAI),\\
Tsukuba, Ibaraki 305-0801, Japan \\
\llap{$^d$}Graduate School of Pure and Applied Science,
University of Tsukuba,\\
Tsukuba, Ibaraki 305-8571, Japan\\
\email{haoki@cc.saga-u.ac.jp,
jnishi@post.kek.jp,
susaki@post.kek.jp}} 

\abstract{
We investigate the effects of non-commutative geometry
on the topological aspects of gauge theory
using a non-perturbative formulation based on the twisted reduced
model.
The configuration space is decomposed into topological sectors
labeled by the index $\nu$ of the overlap Dirac operator 
satisfying the Ginsparg-Wilson relation.
We study the probability distribution of $\nu$ by Monte Carlo 
simulation
of the U(1) gauge theory on 2d non-commutative space with
periodic boundary conditions.
In general the distribution is asymmetric under 
$\nu \mapsto -\nu$,
reflecting the parity violation due to non-commutative geometry.
In the continuum and infinite-volume limits, however,
the distribution turns out to be dominated 
by the topologically trivial sector.
This conclusion is consistent with the
instanton calculus in the continuum theory.
However, it is in striking contrast to the known results in the
commutative case obtained from lattice simulation,
where
the distribution is Gaussian in a finite volume,
but the width diverges
in the infinite-volume limit.
We also calculate
the average action in each topological sector,
and provide deeper understanding of the observed phenomenon.
}


\begin{document}

\section{Introduction}

Non-commutative (NC) geometry \cite{Sny,Connes}
has been studied for quite a long time
as a simple modification of our notion of space-time
at short distances possibly due to effects of 
quantum gravity \cite{gravity}.
It has attracted much attention since it was shown to 
appear naturally from matrix models \cite{CDS,AIIKKT}
and string theories \cite{String}.
In particular, field theory on NC geometry has
a peculiar property known as the UV/IR mixing \cite{rf:MRS},
which may cause a drastic change of the long-distance physics
through quantum effects.
This phenomenon has been first discovered in perturbation theory, 
but it was shown to appear also in a fully nonperturbative
setup \cite{AMNS}.
A typical example is the spontaneous breaking of the
translational symmetry in NC scalar field theory, 
which was first conjectured 
from a self-consistent one-loop analysis \cite{GuSo}
and confirmed later on by Monte Carlo simulation 
\cite{Procs,AC,Bietenholz:2004xs}. (See also \cite{ChenWu,CZ}.)

The appearance of a new type of IR divergence due to the UV/IR mixing
spoils the perturbative renormalizability in most cases 
\cite{Chepelev:1999tt},
and therefore, even the existence of a sensible field theory on a NC
geometry is {\em a priori} debatable.
In order to study such a nonperturbative issue,
one has to define a regularized field theory on NC geometry.
This can be done by using matrix models.
In the case of NC torus, for instance, 
the so-called twisted reduced model \cite{EK,GAO}
is interpreted
as a lattice formulation of NC field theories \cite{AMNS},
in which finite $N$ matrices are mapped
one-to-one onto fields on a periodic lattice.
The existence of a sensible continuum limit
and hence the nonperturbative renormalizability
have been shown by Monte Carlo simulations
in NC U(1) gauge theory in 2d \cite{2dU1} and 4d
\cite{4dU1}
as well as in NC scalar field theory 
in 3d \cite{Bietenholz:2004xs,Bietenholz:2004as}.

In the case of fuzzy sphere \cite{Madore}, finite $N$ matrices
are mapped one-to-one onto functions on the sphere
with a specific cutoff on the angular momentum.
The fuzzy sphere (or fuzzy manifolds 
\cite{Dolan:2003kq,O'Connor:2003aj} in general)
preserves the continuous
symmetry of the base manifold, which makes it an interesting 
candidate for a novel regularization of {\em commutative} field theories
alternative to the lattice \cite{Grosse:1995ar}.
Stability of fuzzy manifolds in
matrix models with the Chern-Simons term \cite{Myers:1999ps,0101102}
has been studied by Monte Carlo simulations
\cite{Azuma:2004zq,fuzzy-MC}.

One of the interesting features of NC field theories
is the appearance of a new type of topological objects,
which are referred to as NC solitons \cite{GMS},
NC monopoles, NC instantons, and fluxons \cite{fluxon}
in the literature.
They are constructed by using a projection operator,
and the matrices describing such configurations
are assumed to be infinite dimensional.
In finite NC geometries topological objects have been constructed 
by using the algebraic K-theory and projective modules 
\cite{non-trivial_config,
Balachandran:2003ay,AIN3}.

Dynamical aspects of these topological objects are
of particular importance
in the realization of a chiral gauge theory in our four-dimensional
world by compactifying a string theory
with a nontrivial index in the compactified dimensions.
Ultimately we hope to realize such a scenario {\em dynamically},
for instance, in the IIB matrix model \cite{9612115},
in which the dynamical generation of {\em four-dimensional} space-time
\cite{Aoki:1998vn,spaceiib,%
gaussian}
as well as the gauge group \cite{Iso:1999xs,0504217}
has been studied intensively.
A crucial link in generating chiral fermions from a matrix model
is provided 
by the index theorem \cite{Atiyah:1971rm}, which relates 
the topological charge of an arbitrary gauge configuration
to the index of the Dirac operator on that background.
The index theorem can be proved 
in noncommutative $\R ^ d$ in the same way as in the
commutative case \cite{Kim:2002qm}.

Extension of the index theorem to {\em finite} NC geometry
is a non-trivial
issue due to the doubling problem of the naive Dirac action.
In lattice gauge theory, an analogous problem was
solved by the use of the so-called overlap 
Dirac operator \cite{Neuberger,Narayanan:1994gw,Hasenfratzindex,Luscher},
which satisfies
the Ginsparg-Wilson relation \cite{GinspargWilson}.
The ideas developed in lattice gauge theory
have been successfully extended to NC geometry.
In the case of NC torus, the overlap Dirac operator has been
introduced in ref.\ \cite{Nishimura:2001dq}, and 
it was used to define a NC chiral gauge theory
with manifest star-gauge invariance.
For general NC manifolds,
a prescription to define 
the Ginsparg-Wilson Dirac operator and its index
has been provided in ref.\ \cite{AIN2},
and the fuzzy sphere was considered as a concrete example
\footnote{The Ginsparg-Wilson Dirac operator for vanishing gauge field 
was constructed earlier in refs.\ \cite{balagovi}.}.
The Ginsparg-Wilson algebra
for the fuzzy sphere 
has been studied in detail in each topological 
sector \cite{Balachandran:2003ay}.
%
In ref.\ \cite{isonagao}
the overlap Dirac operator on the NC torus 
\cite{Nishimura:2001dq}
was derived also from this general prescription \cite{AIN2},
and the axial anomaly has been calculated in the continuum limit.

In an attempt to construct a topologically nontrivial configuration
on the fuzzy sphere, an analogue of the 't Hooft-Polyakov monopole
was obtained \cite{Balachandran:2003ay,AIN3}.
Although the index defined through the Ginsparg-Wilson Dirac operator
vanishes for these configurations, one can 
make it non-zero
by inserting a projection operator, which
picks up the unbroken U(1) component of the SU(2) gauge group.
In fact the 't Hooft-Polyakov monopole configurations are precisely
the meta-stable states observed in Monte Carlo simulations \cite{Azuma:2004zq}
taking the two coincident fuzzy spheres as the initial configuration,
which eventually decays into a single fuzzy sphere.
In ref.\ \cite{AIMN}
this instability was studied analytically by
the one-loop calculation of free energy
around the 't Hooft-Polyakov monopole 
configurations,
and it was interpreted as
the dynamical generation of a nontrivial index,
which may be used for the realization of a chiral fermion in our space-time.
%

In our previous work \cite{Aoki:2006sb}, 
we have demonstrated the validity of 
the index theorem in finite NC geometry,
taking the 2d U(1) gauge theory on a discretized NC torus as a 
simple example,
which is studied extensively in the literature
both numerically \cite{2dU1}
and analytically \cite{Mafia,Paniak:2002fi,
Griguolo:2003kq}.
In particular, ref.\ \cite{Griguolo:2003kq} presents
general classical solutions carrying the topological charge.
We computed the index defined through the Ginsparg-Wilson Dirac operator
for these classical solutions 
and compared the results with the topological charge. 
The index theorem holds when the action is small,
but the index takes only multiple integer values 
of $N$, the size of the 2d lattice.
For non-zero indices, the action is finite in the large $N$ limit,
but it diverges when the bare coupling constant is tuned in the continuum
limit.
By interpolating the classical solutions, we constructed explicit 
configurations
for which the index is of order 1, but
the action becomes of order $N$.
These results 
suggested that the probability of obtaining 
a non-zero index
vanishes
in the continuum limit.

In this paper we confirm this statement 
{\em at the quantum level}
by performing Monte Carlo simulation of
the 2d U(1) gauge theory on a NC discretized torus.
Since the theory is known to have a sensible 
continuum limit \cite{2dU1},
we investigate the probability distribution of the index
in that limit.
Comparison with the known results in the corresponding commutative case
obtained from lattice simulation
\cite{GHL} allows us to reveal the striking effects of NC geometry.


The rest of this paper is organized as follows.
In section \ref{model} 
we define the model and the index of the
overlap Dirac operator.
In section \ref{sec:prob-index} we show our results
for the probability distribution of the index.
In section \ref{sec:average-action} we discuss
the average action in each topological sector,
which provides qualitative understanding for
the behavior of the probability distribution.
Section \ref{summary} is devoted to a summary and discussions.


\section{The model and the topological sectors}
\label{model}
In this section we define the model and 
the topological sectors based on the matrix model formulation
of NC gauge theory.
For more details such as the interpretation
of matrices as fields on a NC torus,
we refer the reader to our previous paper \cite{Aoki:2006sb}.

The model we study in this paper is given by the action
\beq
S = N^2 \beta \, \sum_{\mu \ne \nu} 
\left\{ 1 - \frac{1}{N}
{\cal  Z}_{\nu\mu} \, 
\tr \Bigl(V_\mu\,V_\nu\,V_\mu^\dag\,V_\nu^\dag\Bigr)
\right\}  \ ,
\label{TEK-action}
\eeq
where ${\cal Z}_{\mu\nu}={\cal  Z}_{\nu\mu}^*$ is a
phase factor given by \cite{Nishimura:2001dq}
\beq
{\cal  Z}_{12}=\exp\left(\pi i \frac{N+1}{N} \right)
\label{def-twist}
\eeq
with $N$ being an odd integer.
The NC tensor $\Theta_{\mu\nu}$, 
which characterizes NC geometry
$[x_\mu , x_\nu] = i \Theta_{\mu\nu}$, is given by
\beq
\Theta_{\mu\nu} = \vartheta \, \epsilon_{\mu\nu}  \ ,
\quad \quad \quad
\vartheta =\frac{1}{\pi} N a^2  \ .
\label{theta-def}
\eeq
%



Since the NC parameter $\vartheta$
is related to the lattice spacing by (\ref{theta-def}),
we have to take the large $N$ limit together with the continuum limit
$a \rightarrow 0$ in order to obtain
a continuum theory with finite $\vartheta$.
In that limit the physical extent of the torus $\ell = a N$
goes to infinity at the same time.
Whether one can obtain a sensible continuum limit
by tuning $\beta$ appropriately
is a non-trivial issue, which has been addressed in ref.\ \cite{2dU1}.
It turned out that $\beta$ should be sent to $\infty$ as
\beq
\beta \propto \frac{1}{a^2} \ .
\label{relation_a_beta}
\eeq
Combining this with (\ref{theta-def}), one finds that
the large $N$ limit should be taken together with 
$\beta \rightarrow \infty$ limit so that 
$\beta/N$ is fixed. This limit is
called the ``double scaling limit'', in which non-planar
diagrams survive.
If one takes the planar limit ($N \rightarrow \infty$ with fixed $\beta$)
instead, one obtains a gauge theory 
on a NC space with $\vartheta=\infty$.
In this limit the Wilson loops agree 
\footnote{
At finite $\vartheta$,
the agreement holds 
only when the physical area surrounded by the Wilson loop is much
smaller than $\vartheta$. 
As a consequence, the relation (\ref{relation_a_beta}) agrees
with the one required for the continuum limit of
the SU($\infty$) lattice gauge theory \cite{GW}.
}
with the SU($\infty$) lattice gauge theory \cite{GW}
due to the Eguchi-Kawai equivalence \cite{EK}.
In particular the expectation value of the action in this limit
is given by \cite{GW}
\beq
\langle S \rangle =\left\{
\begin{array}{ccc}
2 \beta N^2 (1-\beta) &~~ \mbox{for} & \beta  < \frac{1}{2}  \ , \\
 \frac{1}{2}N^2 &~~\mbox{for}& \beta \geq \frac{1}{2}  \ ,
\end{array}\right. 
\label{S-GW}
\eeq
which shows that the system undergoes a third order phase transition
at $\beta = \beta_{\rm cr}\equiv 1/2$.


The configuration space can be naturally 
decomposed into topological
sectors by the index of the overlap Dirac operator 
on the discretized NC torus \cite{Nishimura:2001dq,AIN2,isonagao}.
Let us define the covariant forward (backward) difference operator
\beqa
\nabla_\mu \Psi&=&
\frac{1}{a}\left[V_\mu \Psi \Gamma_\mu
- \Psi  \right] \ , \nonumber \\
\nabla_\mu^* \Psi &=&
\frac{1}{a}\left[\Psi - V_\mu ^\dagger 
\Psi  \Gamma_\mu \right] \ ,
\label{def-cov-shift}
\eeqa
where the SU($N$) matrices $\Gamma_\mu$ ($\mu = 1, 2$)
satisfy the 't Hooft-Weyl algebra
\beq
\label{tH-W-alg}
\Gamma_\mu \Gamma_\nu = {\cal  Z}_{\mu\nu}
\Gamma_\nu \Gamma_\mu   \ .
\eeq
Given the covariant forward (backward) difference operator,
we can define the overlap Dirac operator in precisely the same
way as in the commutative case.

First the Wilson-Dirac operator can be defined as
\beq
D_{\rm W}=\frac{1}{2}\sum_{\mu=1}^2
\left\{\gamma_\mu\left(\nabla_\mu^* 
+\nabla_\mu \right) - a \nabla_\mu^* \nabla_\mu \right\} \ ,
\label{def-Wilson-Dirac}
\eeq
where $\gamma_\mu$ ($\mu = 1,2$) are the gamma matrices in 2d.
A crucial property of the overlap Dirac operator $D$
is the Ginsparg-Wilson relation \cite{GinspargWilson}
\beq
\gamma_5 D + D\gamma_5 = a \, D\gamma_5 D \ ,
\label{GWrel}
\eeq
where $\gamma_5 = - i \gamma_1 \gamma_2$ is the chirality operator.
Assuming the $\gamma_5$-hermiticity $D^\dagger =\gamma_5 D\gamma_5$,
we can define a hermitean operator $\hat\gamma_5$ by
\beq
{\hat \gamma_5}=\gamma_5\left(1- a D\right) \ ,
\eeq
which may be solved for $D$ as 
$D=\frac{1}{a} (1 - \gamma_5 \hat{\gamma}_5)$.
Then the Ginsparg-Wilson relation (\ref{GWrel}) is equivalent to
requiring $\hat\gamma_5$ to be unitary.
The overlap Dirac operator 
corresponds to taking $\hat\gamma_5$ to be \cite{Neuberger}
\beqa
\hat\gamma_5 &=& \frac{H}{\sqrt{H^2}} \ , \\
H &=& \gamma_5 \left(1-aD_{\rm W}\right) \ ,
\label{H-def}
\eeqa
where $D_{\rm W}$ is the Wilson-Dirac operator.

One can define the index of $D$ unambiguously by
$\nu \equiv
n_+ - n_-$,
where $n_{\pm}$ is the number of zero modes
with the chirality $\pm 1$.
It turns out that \cite{Narayanan:1994gw,Hasenfratzindex,Luscher}
\beq
\nu 
=\frac{1}{2} {\cal T}r ( \gamma_5 + \hat\gamma_5) 
= \frac{1}{2} {\cal T}r \frac{H}{\sqrt{H^2}} \ , 
\label{indexD-H}
\eeq
where ${\cal T}r$ represents a trace over the space of matrices
and over the spinor index.

We performed Monte Carlo simulation of the model (\ref{TEK-action})
using the heat bath algorithm as in ref.\ \cite{2dU1}.
For each configuration $V_\mu$ generated by simulation,
we calculate the index (\ref{indexD-H}).
We diagonalize the hermitean matrix $H$ defined by (\ref{H-def}), 
and count the number of positive and negative eigenvalues.
(Note that the lattice spacing $a$ which appears in the expressions 
(\ref{def-cov-shift}), (\ref{def-Wilson-Dirac}) and (\ref{H-def})
cancel each other, and the index does not depend explicitly
on $a$.)
The computational effort for calculating
the index is of order $N^6$,
since we have to diagonalize the $2N^2 \times 2N^2$
hermitean matrix $H$.





\section{Probability distribution of the index}
\label{sec:prob-index}

In this section we present our results for 
the probability distribution of the index $\nu$
--- as computed 
by the definition (\ref{indexD-H}) ---
in the gauge theory on the NC torus.

    \FIGURE{
    \epsfig{file=eps_%
Ndistribution_indexoverN_planar_b0.00.eps,
angle=270,width=7.4cm}
\caption{The probability
distribution of $\frac{\nu}{N}$ for various $N$ at $\beta=0$.}
\label{b0}
}

In figure \ref{b0}
we plot the probability distribution of $\nu$ 
for various $N$ at $\beta=0$.
This represents the distribution
in the configuration space
without taking account of the action.
To our surprise, it turns out that the distribution 
of $\nu$ is asymmetric under $\nu \mapsto -\nu$.
This is in striking contrast to ordinary commutative theories,
in which the distribution of $\nu$ is symmetric 
due to parity invariance.
We also find that the distribution for the rescaled topological
charge $\nu/N$ at different $N$ lies on top of each other.
This behavior is analogous to what one obtains in the 
commutative continuum theory (See, for instance,
section 6 of ref.\ \cite{Aoki:2006sb}.).
The plot also confirms 
the existence of $\nu \ne 0$ configurations
on the discretized NC torus. An example of such configurations
is found numerically in ref.\ \cite{Nagao:2005st},
and constructed analytically in 
section 5 of ref.\ \cite{Aoki:2006sb}.
The crucial question we address in what follows is 
whether such configurations survive in the continuum limit.

%


    \FIGURE{
    \epsfig{file=eps_%
distribution_index_N15.eps,%
angle=270,width=7.4cm}
    \epsfig{file=eps_%
distribution_index_planar_b0.55.eps,%
angle=270,width=7.4cm}
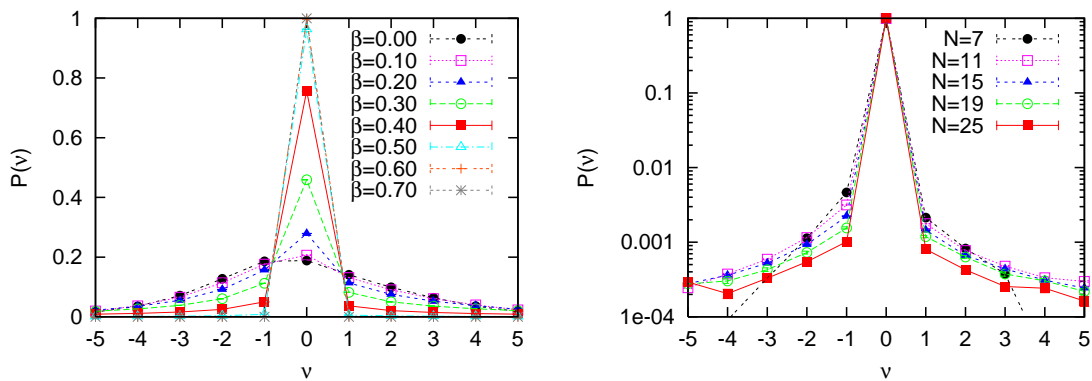
\caption{The probability distribution of $\nu$ is plotted
for various $\beta$ at $N=15$ (left) and 
for various $N$ at $\beta=0.55$ (right).
In the right plot, the log scale is taken for the y-axis to
magnify the results at $\nu \neq 0$.
}
\label{distrib-N-beta}
}

Let us see how 
the probability distribution of $\nu$ changes as we switch on $\beta$.
In figure \ref{distrib-N-beta} (left) we 
plot the probability distribution $P(\nu)$
for various $\beta$ at $N=15$.
(Throughout this paper, we assume the normalization
$\sum_\nu P(\nu) = 1$.)
We find that the probability for $\nu \neq 0$ decreases rapidly,
and the probability for $\nu = 0$ approaches unity.
In figure \ref{distrib-N-beta} (right) we plot 
the probability distribution $P(\nu)$
for various $N$ at $\beta = 0.55$.
Note that the value of $\beta$ we have chosen
is above the critical point $\beta = \beta_{\rm cr}\equiv 1/2$
of the Gross-Witten phase transition.
We find that the distribution approaches
the Kronecker delta $\delta_{\nu 0}$
not only for increasing $\beta$ but also for increasing $N$.
In figure \ref{logrho1overrho0}
we plot the ratio $P(\nu)/P(0)$ for $\nu=1,-1$
for various $\beta$ at $N=15$ (left) and 
for various $N$ at $\beta = 0.55$ (right).
In both cases we observe an exponentially decreasing behavior.

    \FIGURE{
    \epsfig{file=eps_%
rho1overrho0_vs_beta.eps,%
angle=270,width=7.4cm}
    \epsfig{file=eps_%
rho1overrho0_vs_N_b0.55_planar.eps,%
angle=270,width=7.4cm}
\caption{
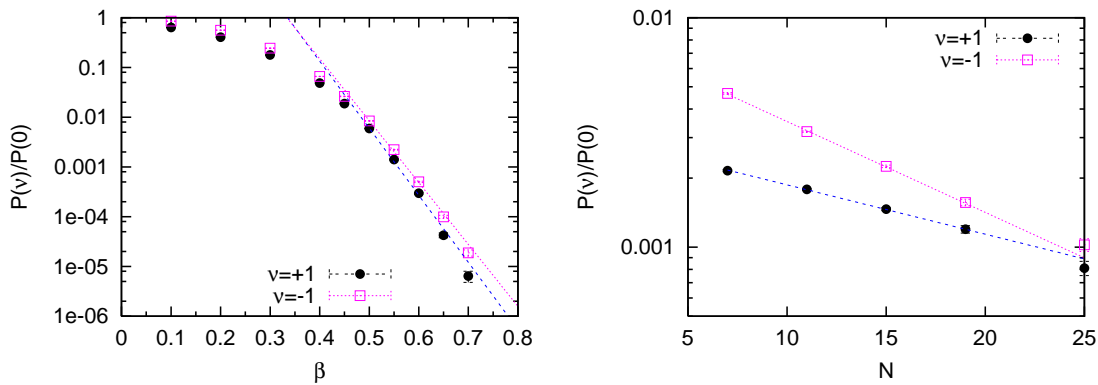
The ratio $P(\nu)/P(0)$ for $\nu=1,-1$
is plotted in the log scale
as a function of $\beta$ at $N=15$ (left) and
as a function of $N$ at $\beta=0.55$ (right).
The straight lines represent a fit to an exponentially decreasing
behavior.
}
\label{logrho1overrho0}
}

As we mentioned in the previous section,
in order to take the continuum limit,
we have to send $N$ and $\beta$ to $\infty$
simultaneously fixing the ratio
$\beta/N$.
It is clear from the above results that
the distribution $P(\nu)$ approaches $\delta_{\nu 0}$
very rapidly in that limit.

\section{Average action in each topological sector}
\label{sec:average-action}

    \FIGURE{
    \epsfig{file=eps_%
distribution_action_b0.10_0.50.eps,%
angle=270,width=7.4cm}
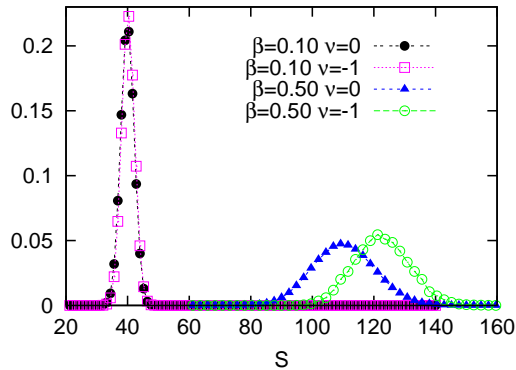
\caption{The distribution of the action in the
$\nu = 0$ and $\nu =-1$ topological sectors
is plotted for $\beta= 0.1$ and $\beta=0.5$ at $N=15$.
}
\label{dist_actionN15}
}

In this section 
we provide an explanation of our results in the previous section
by studying the action in each topological sector.
In figure \ref{dist_actionN15} we plot 
the distribution of the action $S$
for $\nu=0, -1$ at $\beta = 0.1$ and $\beta = 0.5$.
We find that at $\beta = 0.1$ the distribution for different
topological sector lies on top of each other, while
at $\beta = 0.5$ the distribution for $\nu =0$ differs much
from $\nu=-1$. We have also measured 
the distribution for $\nu =1,2,-2$, which turns out to be 
very close to the result for $\nu=-1$.

In figure \ref{actionN15}
we plot the average value of the action $\bar{S}(\nu)$ in 
each topological sector.
We find that the result is almost flat
except at $\nu = 0$.
Note that the weighted sum of $\bar{S}(\nu)$ yields
\beq
\sum_{\nu} \bar{S}(\nu)  P(\nu) = 
\langle S \rangle \ ,
\eeq
where $\langle S \rangle$ is given by (\ref{S-GW})
in the planar limit.
When the $\nu=0$ sector dominates,
we have $\bar{S}(0) \simeq \langle S \rangle$.
This explains the behavior of $\bar{S}(0)$ in figure \ref{actionN15}.

In both plots in figure \ref{actionN15},
we observe a dip at $\nu=0$.
In figure \ref{dipN15}
we plot the size of the dip defined by
\beq
\Delta S \equiv {\bar S}(-1) - {\bar S}(0) \ ,
\eeq
which shows that the dip grows linearly with both $\beta$ and $N$.
(From the left plot, we find that the linear behavior
sets in at $\beta \sim 0.5$, which is close to the critical point
of the Gross-Witten phase transition.)
This is consistent with the exponentially decreasing behavior
of the probability $P(\nu)/P(0)$
for $\nu \neq 0$ discussed in the previous section. 

In the commutative case \cite{GHL},
lattice simulation shows that 
the average action in each sector
increases quadratically with $\nu$, 
but the coefficient vanishes
in the infinite-volume limit.
Correspondingly the distribution of $\nu$
is Gaussian in a finite volume, but the 
width diverges in the infinite-volume limit.
Thus the situation in the NC case differs
drastically from the commutative case.

    \FIGURE{
    \epsfig{file=eps_%
action_vs_index_N15.eps,%
angle=270,width=7.4cm}
    \epsfig{file=eps_%
action_vs_index_planar_b0.55.eps,%
angle=270,width=7.4cm}
\caption{
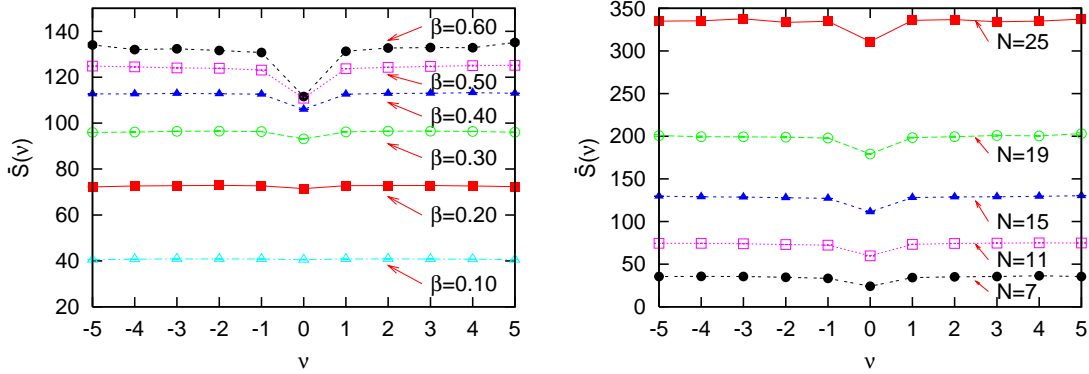
The average value of the action
is plotted 
against the index $\nu$ for various $\beta$ at $N=15$ (left)
and for various $N$ at $\beta=0.55$ (right).
}
\label{actionN15}
}

    \FIGURE{
    \epsfig{file=eps_%
dip_of_nu0andnu-1_vs_beta_N15.eps,%
angle=270,width=7.4cm}
    \epsfig{file=eps_%
dip_vs_N_of_nu0andnu1_planar.eps,%
angle=270,width=7.4cm}
\caption{
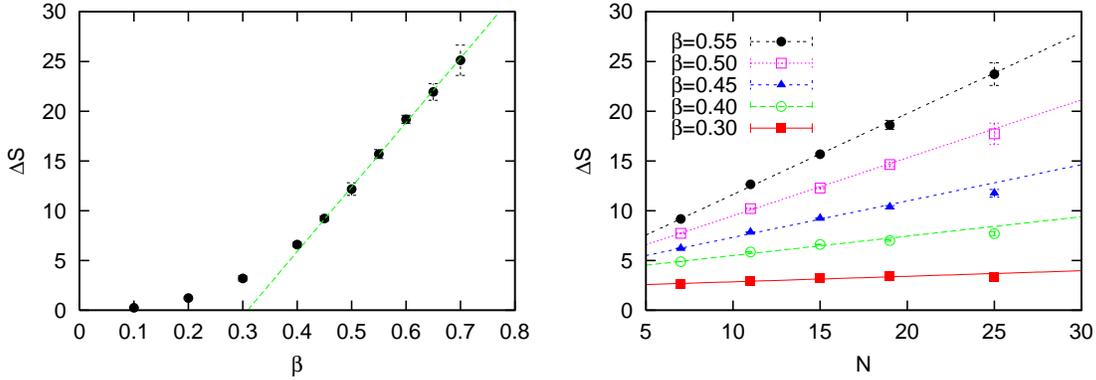
The dip $\Delta S$ is plotted 
as a function of $\beta$ for $N=15$ (left)
and 
as a function of $N$ for various $\beta$ (right).
}
\label{dipN15}
}

\section{Summary and discussions}
\label{summary}

In this paper we have studied the effects
of NC geometry on the probability distribution 
of the index $\nu$ of the Dirac operator.
In the 2d U(1) gauge theory with periodic boundary conditions,
we found that
the probability for $\nu \neq 0$ is exponentially suppressed
in the continuum and infinite-volume limits.
Our conclusion is consistent with our previous analysis at
the classical level \cite{Aoki:2006sb}
and with the instanton calculus
in the continuum theory \cite{Paniak:2002fi}.
In fact the topologically trivial sector includes
all the instanton configurations that contribute to 
the partition function.

In order to understand our conclusion intuitively, 
let us recall that in NC geometry,
the action involves the star product, 
which must have certain smoothening effects 
on the gauge field.
In the commutative case with periodic boundary
conditions,
a classical solution in a topologically non-trivial sector
has a constant field strength, but the vector potential
has a singularity. (See e.g., section 6 of 
ref.\ \cite{Aoki:2006sb}.)
It is therefore conceivable that such configurations 
cannot be realized in NC geometry.
Our results in section \ref{sec:average-action}
substantiate this argument.

It follows from our conclusion that
special care must be taken when
one studies the $\theta$-vacuum in NC
geometry.\footnote{Strictly speaking, we need
to have the ordinary (commutative) time
in order to be able to speak about a ``vacuum''.
We may think of, for instance,
four-dimensional space-time
with non-commutativity introduced only
in two spatial directions \cite{4dU1}.
Let us also remind the readers that
the parameter $\theta$ should not be confused with
the non-commutativity parameter $\vartheta$.}
In general one has to sum over (topologically different) 
twisted boundary conditions labeled by $\nu$ 
with the phase factor $\ee^{i \nu \theta}$.
In the commutative case, however, one may equivalently
add a $\theta$-term to the action and just integrate over
the lattice configuration with periodic 
boundary conditions, as is done e.g., 
in ref.\ \cite{FukayaOnogi}.
Our conclusion implies that this is no longer true
in NC geometry. In ref.\ \cite{Aoki:2006sb}
we speculated that NC geometry may 
provide a solution to the strong CP problem,
but this remains to be seen.

As another effect of NC geometry,
we found that in general the probability distribution 
of 
$\nu$ becomes 
asymmetric under $\nu \mapsto -\nu$,
reflecting the parity violation due to NC geometry.
This is interesting since it suggests
a possibility to obtain a non-zero vacuum expectation value
for the index $\nu$ in some NC model.
Alternatively, one can twist the boundary condition
to make a topologically non-trivial 
sector dominate in the continuum and infinite-volume 
limits \cite{ANS}.
We expect that these unusual properties of NC geometry
may provide
a dynamical mechanism for realizing chiral fermions in
string theory compactifications, or a mechanism for
generating non-zero baryon number density in the universe.
See refs.\ \cite{AIMN,Aschieri:2003vy} for a related line of research
using fuzzy spheres in the extra dimensions.

From the motivations mentioned above, 
it would be interesting to extend the present analysis
to four dimensions.
Unlike the 2d case studied in this paper,
the perturbative vacuum is actually unstable 
due to the UV/IR mixing 
\cite{LLT,ruiz,Bassetto:2001vf,rf:MVR,rf:AL,Guralnik:2002ru}.
However, the system stabilizes after the condensation of the Wilson lines
and finds a stable nonperturbative vacuum
\cite{4dU1},
in which the translational invariance is spontaneously broken.
One can also stabilize the perturbative vacuum by keeping the
UV cutoff finite and regarding the model as a low-energy effective
theory. The situation may depend on which vacuum one chooses.
We hope to address such issues
in future publications.


\acknowledgments

It is our pleasure to thank Hidenori Fukaya,
Satoshi Iso, Hikaru Kawai, Kenji Ogawa and Kentaroh Yoshida
for valuable discussions.

\end{document}